\documentclass[%
reprint,
 superscriptaddress,
 preprintnumbers,
 nofootinbib,
 amsmath,amssymb,
 aps,
 prl,
longbibliography
]{revtex4-1}
\usepackage{psfrag,slashed,cancel,array,graphicx,todonotes,hyperref}
\usepackage{subfigure}
\usepackage[utf8]{inputenc}
\usepackage{mathtools}
\usepackage{mathrsfs}
\usepackage{multirow}
\usepackage{braket}
\usepackage{booktabs} 
\usepackage[normalem]{ulem}
\hypersetup{pdftitle={},pdfcreator={},linkcolor=[rgb]{0.15,0.35,0.75},colorlinks=true,citecolor=[rgb]{0.675,0,0.2},urlcolor=[rgb]{0.15,0.35,0.65}}

\def\beq{\begin{equation}}
\def\eeq{\end{equation}}
\def\bsp#1\esp{\begin{split}#1\end{split}}

\AtBeginDocument{
\heavyrulewidth=.08em
\lightrulewidth=.05em
\cmidrulewidth=.03em
\belowrulesep=.65ex
\belowbottomsep=0pt
\aboverulesep=.4ex
\abovetopsep=0pt
\cmidrulesep=\doublerulesep
\cmidrulekern=.5em
\defaultaddspace=.5em
}

\def\cC{\mathcal{C}}

\newcommand{\rd}{\textrm{d}}

\makeatletter
\def\mr@ignsp#1 {\ifx\:#1\@empty\else #1\expandafter\mr@ignsp\fi}%
\newcommand{\multiref}[1]{\begingroup
\xdef\mr@no@sparg{\expandafter\mr@ignsp#1 \: }%
\def\mr@comma{}%
\@for\mr@refs:=\mr@no@sparg\do{\mr@comma\def\mr@comma{,}\ref{\mr@refs}}%
\endgroup}
\makeatother

\newcommand{\figref}[1]{Fig.~\multiref{#1}}
\renewcommand{\eqref}[1]{(\multiref{#1})}

\DeclareMathOperator{\K}{K}
\DeclareMathOperator{\xr}{cr}

\begin{document}

\preprint{BONN-TH-2022-19}

\title{Yangian-invariant fishnet integrals in 2 dimensions as volumes of Calabi-Yau varieties 
}

\author{Claude Duhr}
\email{cduhr@uni-bonn.de}
\affiliation{Bethe Center for Theoretical Physics, Universit\"at Bonn, D-53115, Germany.}

\author{Albrecht Klemm}
\email{aoklemm@uni-bonn.de}
\affiliation{Bethe Center for Theoretical Physics, Universit\"at Bonn, D-53115, Germany.}
\affiliation{Hausdorff Center for Mathematics, Universität Bonn, D-53115, Germany.}

\author{Florian Loebbert}
\email{loebbert@uni-bonn.de}
\affiliation{Bethe Center for Theoretical Physics, Universit\"at Bonn, D-53115, Germany.}

\author{Christoph Nega}
\email{cnega@uni-bonn.de}
\affiliation{Bethe Center for Theoretical Physics, Universit\"at Bonn, D-53115, Germany.}

\author{Franziska Porkert}
\email{fporkert@uni-bonn.de}
\affiliation{Bethe Center for Theoretical Physics, Universit\"at Bonn, D-53115, Germany.}

\begin{abstract}
We argue that $\ell$-loop Yangian-invariant fishnet integrals in 2 dimensions are connected to a family of Calabi-Yau $\ell$-folds. The value of the integral can be computed from the periods of the Calabi-Yau, while the Yangian generators provide its Picard-Fuchs differential ideal. Using mirror symmetry, we can identify the value of the integral as the quantum volume of the mirror Calabi-Yau. We find that, similar to what happens in string theory, for $\ell=1$ and 2 the value of the integral agrees with the classical volume of the mirror, but starting from $\ell=3$, the classical volume gets corrected by instanton contributions. We illustrate these claims on several examples, and we use them to provide for the first time results for 2- and 3-loop Yangian-invariant traintrack integrals in 2 dimensions for arbitrary external kinematics.\end{abstract}

\maketitle

Multi-loop Feynman integrals are the cornerstone of all modern perturbative approaches to quantum field theory (QFT) and a backbone of precision computations for collider and gravitational wave experiments. It is therefore of utmost importance to have efficient ways for their computation and a solid understanding of the underlying mathematics.  Over the last years, it has become clear that the mathematics relevant to Feynman integrals is tightly connected to certain topics in geometry. One of the earliest observations was that 1-loop Feynman integrals compute the volumes of certain polytopes in hyperbolic spaces~\cite{Davydychev:1997wa,Schnetz:2010pd,Mason:2010pg,Spradlin:2011wp,Bourjaily:2019exo}. Here the most prominent example is the 1-loop 4-point function with massless propagators in 4 space-time dimensions:
\beq\label{eq:4-m-box-4D}
\int \frac{\rd^4\xi}{i\pi^2}\,\prod_{i=1}^4\frac{1}{(\xi-\alpha_i)^2}= -\frac{4}{\alpha_{13}^2\alpha_{24}^2}\,\frac{D(z)}{z-\bar{z}}
\eeq
with $\alpha_{ij} = \alpha_i-\alpha_j$. This integral features a so-called \emph{(dual) conformal symmetry}~\cite{Drummond:2006rz} (with conformal weight 1 for each external point), and the variable $z$ is connected to the cross ratio formed out of the 4 external points $\alpha_i$.
The function $D(z)$ is the so-called \emph{Bloch-Wigner dilogarithm}, which is known to compute the volume of a simplex in hyperbolic 3-space, see e.g.~\cite{Zagier2007}:
\beq
D(z) = \textrm{Im}\big[\textrm{Li}_2(z)+\log|z|\,\log(1-z)\big]\,.
\eeq
Notably, this result for the 4-point integral can be bootstrapped from scratch, using a Yangian extension of the (dual) conformal symmetry \cite{Loebbert:2019vcj}.

The interpretation of Feynman integrals as volumes is so far only understood at 
1 loop. While there is substantial evidence that, at least in special QFTs, the integrands of Feynman integrals are related to certain volume forms for generalisations of polytopes (see, e.g.,~\cite{Arkani-Hamed:2013jha,Arkani-Hamed:2017mur,Salvatori:2018aha,Arkani-Hamed:2019vag,Damgaard:2020eox}), it is an open question if at higher loops the values after integration can be interpreted as volumes of geometric objects. If that was indeed the case, it would shed new light on the mathematical structure of perturbative QFT, and possibly lead the way towards novel methods for the computation of Feynman integrals. The main goal of this paper is to take first steps into this direction and to present for the 
first time an infinite class of higher-loop Feynman integrals whose values can indeed be interpreted as a volume.

\begin{figure}[t]
\centering
\includegraphics{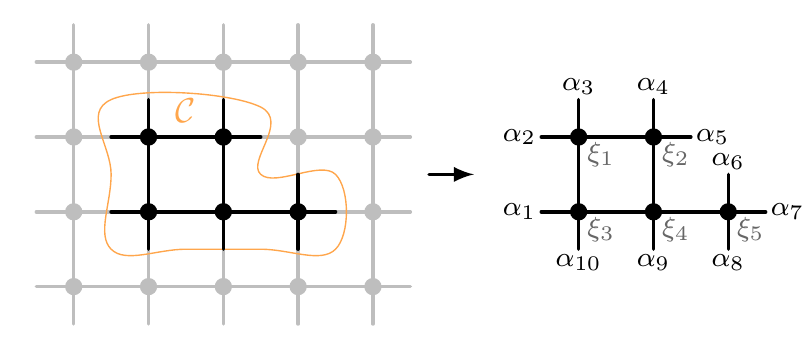}
\caption{\label{fig:fishnet}
Ten-point five-loop fishnet integral cut out of a square tiling of the plane. If the 
$\ell = M\times N$ interior points span a rectangle, we denote the graph by $G_{M,N}$, with $M\le N$.} 
\end{figure}
\section{Fishnet integrals in 2 dimensions}

In the remainder of this paper we focus on so-called 
\emph{fishnet integrals} in 2 Euclidean dimensions, defined by a connected 
region cut out along a closed curve $\cC$ intersecting the edges of a regular tiling of the plane by a square lattice (see \figref{fig:fishnet} and \cite{Chicherin:2017cns}). This defines a connected graph $G$ by considering only the edges of the lattice that intersect $\cC$ (the \emph{external edges}) or lie in its interior (the \emph{interior edges}).
Edges of $G$ connecting 2 vertices labeled by $a,b\in\mathbb{R}^2$ represent propagators $[(a-b)^2]^{-1/2}$, and we integrate over the positions of the internal vertices labeled $\xi_i$. 
It is well known that for a 2-dimensional QFT it is useful to consider complexified coordinates
$a_j = \alpha_j^1 + i\, \alpha_j^2$ and $x_k = \xi_k^1 + i\, \xi_k^2$.
The integrals we want to consider can then be written as 
\beq\label{eq:I_G_complex}
I_G(a) = \int \left(\prod_{j=1}^\ell\frac{\rd x_j\wedge\rd \bar{x}_j}{-2i}\right)\frac{1}{\sqrt{|P_G(x,a)|^2}}\,,
\eeq
with $a= (a_1,\ldots,a_n)$, $x = (x_1,\ldots,x_\ell)$, and
\beq\label{eq:P_G_def}
P_G(x,a) = \Big[\prod_{i,j}(x_i-x_j)\Big]\,\Big[\prod_{i,j}(x_i-a_j)\Big]\,,
\eeq
and the product ranges depend on the graph topology.

Every fishnet integral is invariant under the Yangian $Y(\mathfrak{so}(1,3))$ of the conformal group in 2 dimensions~\cite{Chicherin:2017frs,Loebbert:2020hxk}. The Yangian splits into holomorphic and anti-holomorphic parts:
\beq
Y(\mathfrak{so}(1,3)) = Y(\mathfrak{sl}_2(\mathbb{R})) \oplus \overline{Y(\mathfrak{sl}_2(\mathbb{R}))}\,,
\eeq 
where the generators of $Y(\mathfrak{sl}_2(\mathbb{R}))$ act via partial differential operators on the holomorphic external points $a_i$ and annihilate the integral (at least for generic values of the external points). For the explicit form of the representation of the Yangian, we refer to~\cite{Loebbert:2020hxk}. We note that invariance under the conformal subalgebra $\mathfrak{sl}_2(\mathbb{R})$ implies that we can write $I_G(a) = |F_G(a)|^2\,\phi_G(z)$, where $z=(z_1,\ldots,z_{n-3})$ is a vector of conformal cross ratios formed out of the $a_i$ and $F_G(a)$ is a holomorphic algebraic function that carries the conformal weight. 

Analytic results are known for various classes of fishnet integrals depending only on 4 external points, and consequently only on a single cross ratio, which we choose as $z=\xr(2,3,1,4) = a_{23}a_{14}/a_{21}a_{34}$. In~\cite{Derkachov:2018rot} 
analytic results are given for $G_{M,N}$, where the external points that are incident to the same side of the rectangle are identified (we call these graphs $G_{M,N}^1$) with determinants of (derivatives of) ladder graphs $G_{1,\ell}^1$. The ladder graphs $G_{1,\ell}^1$ themselves can be expressed as bilinear combinations of generalised hypergeometric functions. So far, no results are known for fishnet graphs in 2 dimensions depending on more than 1 cross ratio.

\section{2D fishnets and Calabi-Yau geometries}
We now argue that to every $\ell$-loop fishnet graph we can associate a CY $\ell$-fold. Loosely speaking, a CY $\ell$-fold is a complex $\ell$-dimensional K\"ahler manifold $M_\ell$ that admits a unique holomorphic $(\ell,0)$-form $\Omega$. This last condition can be phrased as follows: the cohomology groups $H^r(M_\ell)$ admit a decomposition
\beq
H^r(M_\ell) = \bigoplus_{p+q=r}H^{p,q}\,,
\eeq
where the $H^{p,q}$ are generated by cohomology classes of $(p,q)$-forms, i.e., forms involving exactly $p$ holomorphic and $q$ anti-holomorphic differentials. The Hodge numbers of $M_\ell$ are $h^{p,q} = \dim H^{p,q}$. The CY condition then translates into $h^{\ell,0} = 1$. Note that for a family of CY varieties parametrised by $d_M$ independent moduli, we have $h^{\ell-1,1}=d_M$  for $\ell\neq 2$. For  K3 families $\ell=2d_M$ is the number of independent transcendental cycles minus two. 

One possibility to define a family of CY $\ell$-folds is given by a double cover. Here we consider the constraint $y^2=P_G(x;a)$,
double covering an $\ell$-dimensional projective base space $B$ with coordinate $x=(x_1,\ldots,x_\ell)$ and canonical class $K_B>0$. This defines a family $M_G$ parametrised by $a$ and with $(\ell,0)$-form 
\beq
\Omega_G = \frac{\mu_B(x)}{\sqrt{P_G(x;a)}}\,,
\label{eq:Omega}
\eeq
where $\mu_B$ is the holomorphic  measure on $B$. Note that $I_G$ is obtained by integrating $\Omega_G\wedge\overline{\Omega}_G$ over $\mathbb{C}^\ell$. To guarantee that we really obtain a family of CY $\ell$-folds, the degree of $P_G(x;a)$ has to be such that the canonical class vanishes. We consider $B=\bigtimes_{i=1}^l\mathbb{P}^1$ and $\mu_B=\wedge_{i=1}^\ell (x'_i {\rm d} x_i-x_i {\rm d} x'_i)$ (with $[x_i:x'_i]$ the homogeneous coordinate on the $i^{\textrm{th}}$ copy of $\mathbb{P}^1$), which is a natural compactification of the integration range $\mathbb C^\ell$ in \eqref{eq:I_G_complex}. The vanishing of the canonical class then translates into the fact that $P_G(x;a)$ has to be of degree 4 in each $\mathbb P^1$. This condition is always fulfilled for fishnet graphs, because all internal vertices are 4-valent. For $\ell>1$, $M_G$ is typically a singular variety.
Similar to~\cite{Klemm:2019dbm}, in all examples that we have studied (see below), these singularities can be resolved by deforming $M_G$ to a smooth CY $\ell$-fold, and we expect this to hold in all cases. We will further elaborate on this in~\cite{CY_paper_2}.

There is a natural set of integrals, called \emph{periods}, that we can associate to a CY $\ell$-fold by integrating $\Omega_G$ over a basis of cycles $\Gamma_i$ that span the middle homology $H_\ell(M_G,\mathbb{Z})$. The vector of periods is 
\beq\label{eq:period_vector}
\Pi_G = \Big(\int_{\Gamma_1}\!\!\!\Omega_G,\ldots,\int_{\Gamma_{b_\ell}}\!\!\!\!\!\!\Omega_G\Big)\,,\,\,\,\,\, b_\ell = \dim H_\ell(M_G,\mathbb{Z})\,.
\eeq
The periods are multivalued functions of $a$.
For every CY $\ell$-fold, there is a monodromy-invariant matrix $\Sigma$ that defines a bilinear pairing on the periods,
and $\Sigma$ may be chosen symmetric for $\ell$ even and anti-symmetric (and even symplectic) for $\ell$ odd.
It is well known how to relate the integral of $\Omega_G\wedge\overline{\Omega}_G$ to the monodomy-invariant combination of periods $\Pi_G^{\dagger}\Sigma\Pi_G$. This gives us a way to reduce the computation of fishnet integrals to the problem of finding the periods of $M_G$. The periods are solutions to certain differential equations, as we will now review.

The flatness of the Gauss-Manin connection implies the existence of an ideal of differential operators, called the \emph{Picard-Fuchs differential ideal (PFI)}, whose space of solutions is precisely spanned by the periods. The PFI can be derived by the Griffiths reduction 
   method or a reduction of the Gel'fand-Kapranov-Zelevisk\u{\i} system, 
   see \cite{MR3965409} for a review. In practice, these methods can 
   be rather slow, in particular in the case of many variables. We find that the PFI of $M_G$ contains the generators of $Y(\mathfrak{sl}_2(\mathbb{R}))$. Moreover, the group $\mathrm{Aut}(G)$ of automorphisms of $G$ acts on $Y(\mathfrak{sl}_2(\mathbb{R}))$ by permuting the external points $a_i$, and so the PFI naturally also contains these operators. Remarkably, in all cases we have studied, the complete PFI of $M_G$ is obtained in this way! 
   
We can summarise our findings as follows.

\begin{center}\fbox{\begin{minipage}{8cm}{\bf Claim 1:} 
For every $\ell$-loop fishnet graph $G$, there exists a family of CY $\ell$-folds $M_G$ with holomorphic $(\ell,0)$-form $\Omega_G$ such that
\beq\bsp\label{eq:IGCYM}
I_G(a)&\, = (-i)^\ell \, \Pi^{\dagger}_G\,\Sigma\,\Pi_G\,,
\esp\eeq
and the PFI is generated by $\mathrm{Aut}(G)\cdot Y(\mathfrak{sl}_2(\mathbb{R}))$.
\end{minipage} }\end{center}
Let us make some comments about this result.
First, Claim 1 implies that the periods of $M_G$ are Yangian invariants.
The invariance under the conformal $\mathfrak{sl}_2(\mathbb{R})$ subalgebra implies that we can write $\Pi_G(a) = F_G(a)\widetilde{\Pi}_G(z)$, where $F_G(a)$ is a holomorphic and algebraic function of $a$. 
Second, we expect that the PFI has a point of maximal unipotent monodromy (MUM).\footnote{For a monodromy matrix $M$ to be maximal unipotent means  
that $(M-{\bf 1})^{n}=0$ only for $n\ge \ell+1$. This implies the  logarithmic  degeneration of the periods discussed below.} For example for traintrack graphs $G_{1,\ell}$, we find that a MUM point can be identified as follows: we label an external point on a small side by $a_1$, and the others clockwise up to $a_{2\ell+2}$.  Then a 
MUM  point is at  $z=0$, with 
$z_k=\xr(1,k+1,k+2,\ell+2)$, $z_\ell=\xr(1,\ell+1,2\ell+2,\ell+2)$ and $z_{\ell+k}=\xr(1,2\ell+3-k,2\ell+2-k,\ell+2)$ for $k=1,\hdots, \ell-1$.  
For the 1-parameter graphs  $G_{M,N}^1$, we found MUM points up to $\ell=M\times N=12$ and a non-orientable graph,
and we expect that they are present in full generality. Near the MUM-point $z=0$, there is a unique holomorphic period $\widetilde{\Pi}_{G,0}(z)$, 
that we can  normalise  to $\widetilde{\Pi}_{G,0}(z) = 1+ \mathcal{O}(z)$,  and which multiplies the  $d_M$ solutions  linear in the 
logarithm, as well as the solution of maximal order $\ell+1$ in the logarithms.   
We define $\phi_G(z) = (-i)^\ell\,\widetilde{\Pi}^{\dagger}_G\,\Sigma\,\widetilde{\Pi}_G$. 
Finally, it is well known that $\widetilde{\Pi}_G^{\dagger}\,\Sigma\,\widetilde{\Pi}_G$ is proportional to $e^{-K(z,\bar{z})}$, where $K(z,\bar{z})$ is the K\"ahler potential for the Weil-Peterssen metric on the moduli space of $M_G$. This gives an interpretation of the Feynman integral in terms of the geometry. In the next section, we relate it  to the quantum volume of the mirror.

\begin{figure}[!t]\centering
	\includegraphics[scale=.92]{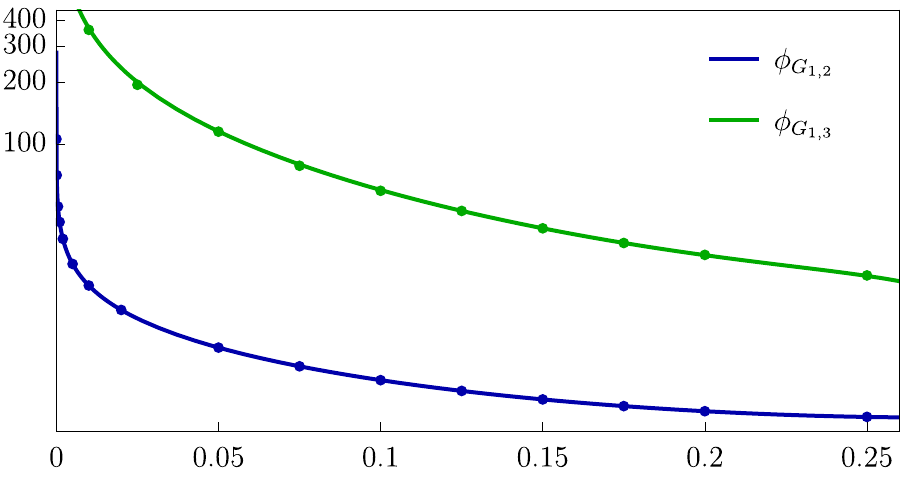}\\[3ex]
\caption{The functions $\phi_{G_{1,2}}(z_1,z_2,z_3)$ and  $\phi_{G_{1,3}}(z_1,\ldots,z_5)$ evaluated on the 1-dimensional slice $(z_1,z_2,z_3) = \tfrac{s}{16}(1,2,3)$ and $(z_1,\ldots,z_5) = \tfrac{s}{16}(1,2,12,4,5)$ (for the definition of our cross ratios, see the main text). The continuous lines represent the results obtained from our analytic result in terms of CY periods, while the dots are obtained from a numerical evaluation of the Feynman parameter representation of $G_{1,\ell}$. }
\label{figure:plots}
\end{figure}

We have verified that we can reproduce the complete PFI from the Yangian generators for $G_{1,2}$, $G_{1,3}$ and $G_{2,2}$.
Having at our disposal the PFI of $M_G$, we can solve the differential equations satisfied by the periods using standard techniques in terms of series expansions~\cite{MR0010757}. This basis of solutions, however, will in general be a linear combination with complex coefficients of the periods in~\eqref{eq:period_vector}. With the methods described in~\cite{Bonisch:2020qmm,Bonisch:2021yfw}, it is possible to construct the change of basis and to find the monodromy invariant bilinear pairing $\Sigma$, and thus to compute $I_G(a)$ through~\eqref{eq:IGCYM}. We have done this explicitly for $G_{1,2}$ and $G_{1,3}$. We have checked that our results numerically agree with a direct evaluation of the Feynman parameter representation for $I_G(a)$ for various values of $a$, and we find very good agreement (see \figref{figure:plots}). 
More details about the structure and the properties of the solutions will be provided in~\cite{CY_paper_2}.

Let us conclude by commenting on the structure of the 4-point ladder graphs $G_{1,\ell}^1$ of~\cite{Derkachov:2018rot,Corcoran:2021gda}. 
At $\ell$ loops we have a 1-parameter family of CY $\ell$-folds whose PFI is generated by a single operator $L_{\ell}$ of degree $\ell+1$ 
that has a MUM-point at $z=0$, and we have\footnote{The  horizontal cohomology $h_{\textrm{hor.}}^{p,q}$ plays  here a similar role as for the Banana 
graphs~\cite{Bonisch:2020qmm,Bonisch:2021yfw}, where the terminology is explained.} $h^{p,\ell-p}_{\textrm{hor.}} = 1$, $0\le p\le \ell$. These operators are special instances of the \emph{Calabi-Yau operators} considered in~\cite{BognerCY,MR3822913}. We have checked up to $\ell=5$ that we reproduce the results of~\cite{Derkachov:2018rot} from our~\eqref{eq:IGCYM}. For $\ell=1$, we obtain the Legendre family of elliptic curves, and the periods can be expressed in terms of elliptic integrals~\cite{Derkachov:2018rot,Corcoran:2021gda}: 
\beq\bsp\label{eq:oneloop}
	\phi_{G_{1,1}^1}\!(z)&\,= \frac{4}{\pi^2}\,(\K\,\overline{\K}'+\overline{\K}\,\K')=\frac{8}{\pi^2}\,|\K|^2\,\textrm{Im}\,\tau_1	\,,
\esp\eeq
and $F_{G_{1,1}^1}\!(a) =\pi/\sqrt{a_{12}a_{34}}$.
Here $\K\equiv\K(z)$ is the complete elliptic integral of the first kind and $\K'\equiv\K(1-z)$, and we defined $\tau_1=i\K'/\K$. 
For $\ell=2$, we obtain a 1-parameter family of K3 surfaces. It is known that every CY operator of degree 3 is equivalent to the symmetric square of a CY operator of degree 2~\cite{doran,BognerCY}, and so we can express $G_{1,2}^1$ in terms of elliptic integrals. We define $\K_\pm = \K\big(\frac12(1\pm\sqrt{1-z})\big)$
such that $(\K_-^2,\K_- \K_+,\K_+^2)$ span the solution space of $L_{3}$. $G_{1,2}^1$ is then given by:
\begin{equation}
\begin{aligned}\label{eq:twoloop}
	\phi_{G_{1,2}^1}\!(z)&=	\frac{2}{\pi^4}	(\K_+  \overline{\K}{}_- + \K_- \overline{\K}{}_+ )^2\\
	&=	\frac{8}{\pi^4}\,|\K_-|^4(\textrm{Im}\,\tau_2)^2 \,,
\end{aligned}
\end{equation}
with $\tau_2=i\K_+/\K_-$ and {$F_{G_{1,2}^1}\!(a) \!=\! 4\sqrt{2}\pi^2/\sqrt{a_{12}a_{34}a_{24}}$}.
For $\ell>2$, it is not possible anymore to express the periods of $M_{G_{1,\ell}^1}$ in terms of elliptic integrals.

\section{Fishnets as quantum volumes}

The results of the previous section allow us to reduce the computation of $I_G(a)$ to the computation of the periods of $M_G$. Since~\eqref{eq:4-m-box-4D} computes the volume of a hyperbolic simplex, it is natural to ask if we can interpret $I_G(a)$ as a volume of sorts. At this point, however, we face an issue. In~\eqref{eq:4-m-box-4D} the ambient hyperbolic space of the simplex provides the canonical metric w.r.t.\ which the volume is computed. On $M_G$, however, we do not have any distinguished metric. Indeed, while $a$ fixes $\Omega_G$, and thus the complex structure, there is still substantial freedom to define a K\"ahler form, and thus a metric, on $M_G$. We now argue that we obtain a volume interpretation using mirror symmetry.

Mirror symmetry expresses the remarkable fact that CY $\ell$-folds come in pairs $(M_G,W_G)$ such that the cohomology groups $H^{p,q}(M_G)$ and $H^{\ell-p,q}(W_G)$ are exchanged. In particular, mirror symmetry exchanges the complex structures encoded in $H^{\ell-1,1}(M_G)$ with the K\"ahler structures from $H^{1,1}(W_G)$. Since $G$ defines via $\Omega_G$ a complex structure on $M_G$, mirror symmetry provides a K\"ahler form $\omega_G\in H^{1,1}(W_G)$. Choosing $z$ such that $M_G$ has a MUM-point at $z=0$, we have
\beq
\omega_G = \sum_i t_{G,i}^{\mathbb{R}}(z)\,\omega^{(i)}\,,
\eeq
where $\omega^{(i)}$ is a basis of $H^{1,1}(W_G)$, and the $t_i^{\mathbb{R}}(z) = \textrm{Im}\,t_i(z)$ are given by the \emph{mirror map} 
\beq
t_{G,i}(z) = \widetilde{\Pi}_{G,i}(z)/\widetilde{\Pi}_{G,0}(z)\,,\quad i=1,\ldots,d_{M}\,,
\eeq where the $\widetilde{\Pi}_{G,i}(z)$ diverge like a single power of a logarithm at the MUM-point.
The K\"ahler form, in turn, can be used to define a volume form $\omega_G^\ell/\ell!$ on $W_G$, and we can define the \emph{classical} volume of $W_G$ as
\beq\bsp\label{volcl}
\textrm{Vol}_{\textrm{cl}}(W_G) &\,= \int_{W_G}\frac{\omega_G^\ell}{\ell!} \\
&\,= \frac{1}{\ell!}\sum_{i_1,\cdots,i_{\ell}}C^{\textrm{cl}}_{i_1,\cdots,i_{\ell}}\,t_{i_1}^{\mathbb{R}}(z)\cdots t_{i_\ell}^{\mathbb{R}}(z)\,,
\esp\eeq
where the $C^{\textrm{cl}}_{i_1,\cdots,i_{\ell}}$ are explicitly-computable integers, namely the (classical) intersection numbers of $M_G$.

Let us illustrate this on the examples of the ladder graphs considered at the end of the previous section. At 1 loop, we find $\textrm{Vol}_{\textrm{cl}}(W_{G_{1,1}^1}) = t_{G_{1,1}^1,1}^{\mathbb{R}}(z) = \textrm{Im}\,\tau_1$, which is the area of the fundamental parallelogram (with sides $(1,\tau_1)$) that defines the elliptic curve $W_{G_{1,1}^1}$ associated to $G_{1,1}^1$. Similarly, we have $\textrm{Vol}_{\textrm{cl}}(W_{G_{1,2}^1}) =\frac{1}{2} t_{G_{1,2}^1,1}^{\mathbb{R}}(z)^2 = \frac{1}{2}(\textrm{Im}\,\tau_2)^2$.
Comparing this to~\eqref{eq:oneloop} and~\eqref{eq:twoloop}, we see that 
\beq\bsp
\label{elk3}
\phi_{G_{1,1}^1}(z) &\,= \frac{4}{\pi^2}\,|\K|^2\,\textrm{Vol}_{\textrm{cl}}(W_{G_{1,1}^1})\,,\\
\phi_{G_{1,2}^1}(z) &\,= \frac{16}{\pi^4}\, |\K_-|^4\,\textrm{Vol}_{\textrm{cl}}(W_{G_{1,2}^1})\,,
\esp\eeq
i.e., we see that the 1- and 2-loop ladder integrals are proportional to the classical volume of the mirror CY 
(the prefactor proportional to $\widetilde{\Pi}_{G_{1,\ell}^1,0}$ defines the overall scale). 
We checked that the same statement holds for the 2-loop traintrack integral $G_{1,2}$, which 
depends on 3 independent cross ratios. However, starting from 
3 loops, the last factor in \eqref{elk3} is no longer proportional to $\textrm{Vol}_{\textrm{cl}}(W_{G})$. 
This is not surprising: it is well known from string theory and mirror symmetry 
that for $\ell>2$ volumes of CY $\ell$-folds receive instanton corrections of order $e^{-t_i^{\mathbb{R}}}$. 
Their contribution is included in the \emph{quantum} volume of $W_G$, 
 \begin{center}\fbox{\begin{minipage}{8cm}{\bf Claim 2:} 
$\phi_G(z)$  is determined by the {quantum} volume of the mirror $W_G$ to $M_G$:
\beq\label{eq:I_G_volume}
\phi_G(z) = |\widetilde{\Pi}_{G,0}(z)|^2\, \textrm{Vol}_\textrm{q}(W_G)\, .
\eeq
\end{minipage} }\end{center}
Note that  one could impose the following requirements on the quantum volume:
i) It is real and positive; ii) it approaches in the limit $z_i\rightarrow 0$ (or equivalently in the
large volume limit $t_i^{\mathbb{R}}\rightarrow \infty$) the classical volume \eqref{volcl}; iii) it is monodromy-invariant, i.e., it uniquely extends over the complex moduli space. 
In \eqref{eq:I_G_volume},  $\textrm{Vol}_\textrm{q}(W_G)$ 
fulfils i) and ii) but not iii).  Because of the normalisation of $\widetilde{\Pi}_{G,0}(z) =1+{\cal O}(z)$ one 
could define $\phi_G$ itself as the quantum volume, fulfilling i)-iii). Nevertheless $\textrm{Vol}_\textrm{q}(W_G)$
seems the more canonical generalisation of the likewise not monodromy-invariant classical 
volumes in \eqref{elk3}. Its CY 3-fold version also features in the analysis of non-perturbative 
properties of string compactifications in \cite{Lee:2019wij}.           

We have checked Claim 2 on our examples for multi-loop traintrack integrals, as well as for the 1-parameter rectangular fishnet integrals of~\cite{Derkachov:2018rot}.
Claim 2 shows that it is possible to give a volume interpretation to all Yangian-invariant fishnet graphs. This extends the volume interpretation of~\eqref{eq:4-m-box-4D} from 4 to 2 dimensions, but with the advantage that the interpretation naturally extends to higher loops. To our knowledge, this is the first time that multi-loop Feynman integrals were identified that compute volumes of geometric objects.

\section{Conclusion}
In this letter we have studied a class of Feynman integrals that connect research in mathematics, scattering amplitudes and integrability. Our main result is that Yangian-invariant $\ell$-loop fishnet integrals in 2 dimensions are naturally associated to families of CY $\ell$-folds and that the values of these integrals represent the quantum volume of the mirror CY. Indeed, we find that for $\ell\le 2$, the traintrack integrals compute the classical volume of the mirror, in agreement with the fact that there are no instanton corrections for elliptic curves or K3 surfaces. Starting from $\ell=3$, instanton corrections can no longer be neglected. This is the first time that it was possible to identify a higher-loop Feynman integral as a volume of a geometric object. Intriguingly, we find that mirror symmetry plays an important role in this context. 

Our results are not just of formal interest. Indeed, we find that we can reduce the problem of computing fishnet integrals in 2 dimensions to the geometrical question of finding the periods of $M_G$. Remarkably, solving this geometrical question receives input from physics, because we find that the Picard-Fuchs differential ideal is determined by the Yangian generators for fishnet graphs (and permutations thereof). We have illustrated this by providing for the first time results for 2-dimensional traintrack integrals at 2 and 3 loops.

Our work opens up several new directions for research, both in mathematics and in physics. First, it would be very interesting to study the geometrical properties of the CY varieties we have encountered in detail, in order to understand what role Yangian symmetry plays from the geometrical point of view. 
It is well known that 1-parameter families of CYs in various dimensions can be related by Hadamard-, symmetric- or anti-symmetric products~\cite{MR3822913}. As an example for 
the last relation we find that the solution space of the 4-point $G_{M,N}^1$ graphs is spanned, possibly up to rational functions~\cite{CY_paper_2}, 
by  $M\times M$ sub-determinants of the Wronskian of the solutions  of the  $G_{1,M+N-1}^1$ graphs.   These determinant relations are reminiscent of 
Basso-Dixon formul\ae~\cite{Basso:2017jwq,Derkachov:2018rot}, but relate integrals of different loop order. From the physics perspective, it would be interesting to clarify the role of the CY geometry in the context of the integrable fishnet theories defined in~\cite{Gurdogan:2015csr,Kazakov:2018qbr}, and in how far the CY geometry, and in particular the instanton corrections for $\ell>2$, play a role in the integrability of the theory. Finally, it would be important to clarify if a similar volume interpretation can also be achieved for other classes of multiloop Feynman integrals, including integrals in 4 space-time dimensions. The most natural place to start is to consider rectangular 4-point fishnet integrals in 4 dimensions, for which analytic results in terms of polylogarithms are known~\cite{Usyukina:1992jd,Usyukina:1993ch,Basso:2017jwq}. Recently, it was shown that $\ell$-loop Yangian-invariant traintrack integrals in 4 dimensions are related to CY $(\ell-1)$-folds~\cite{Bourjaily:2017bsb,Bourjaily:2018ycu,Bourjaily:2018yfy,Bourjaily:2019hmc,Vergu:2020uur} 
(and at 2 loops complete analytic results are known~\cite{Ananthanarayan:2020ncn,Kristensson:2021ani,Wilhelm:2022wow}). It would thus be interesting to investigate if also in this case it is possible to identify a volume description via mirror symmetry.

\begin{acknowledgments}
\emph{Acknowledgements:}  AK likes to thank Sheldon Katz, Maxim Kontsevich,
George Oberdieck and Timo Weigand  for discussions on the geometry realisation of the 
amplitude and  Dr. Max R\"ossler, the Walter Haefner Foundation  as well  the IHES for support. 
FL would like to thank Luke Corcoran and Julian Miczajka for helpful discussions and related collaboration and the \emph{QFT and String Theory Group} at Humboldt University Berlin for hospitality.
The work of FL is funded by the Deutsche Forschungsgemeinschaft (DFG, German Research Foundation)-Projektnummer 363895012, and by funds of the Klaus Tschira Foundation gGmbH. 
\end{acknowledgments}

\bibliography{References}

\end{document}